\documentclass{llncs}
\usepackage{graphicx}

\begin{document}
\pagestyle{empty}
\bibliographystyle{unsrt}
\mainmatter

\title{Cooperative Processes for Scientific Workflows}
\vspace{-0.5cm}
\author{Khaled Gaaloul \and Fran\c cois Charoy \and Claude Godart}
\authorrunning{Khaled Gaaloul et al.}

\institute{
  LORIA - INRIA - CNRS - UMR 7503\\
  BP 239, F-54506 Vand\oe uvre-l\`es-Nancy Cedex, France\\
  \email{\{kgaaloul,charoy,godart\}@loria.fr}
  }
\maketitle

\begin{abstract}
The work described in this paper is a contribution to the problems 
of managing in data-intensive scientific applications. First, we discuss
scientific workflows and motivate there use in scientific applications. 
Then, we introduce the concept of cooperative processes and describe their interactions and uses in a flexible cooperative workflow system called \textit{Bonita}. Finally, we propose an approach to integrate and 
synthesize the data exchanged by the mapping of data-intensive science into Bonita, using a binary approach, and illustrate the endeavors done to enhance the performance computations within a dynamic environment.

{\bf Keywords:} Scientific workflow, data-intensive science, Bonita, PBIO.
\end{abstract}
\vspace{0.5cm}

\section{Introduction}
\label{section:introduction}

With the technological improvements and continuous increasing of scientific requirements, scientific applications are becoming crucial, involving large-scale distributed environments and huge data exchanges. Since data processing can be costly, sharing these data products can save the expense of performing redundant computations \cite{DBLP:conf/eagc/DeelmanBGKMPSVL04}. The goal of scientific workflows is to provide an environment where scientists share their resources, data, applications, and knowledges to pursue common goal. Indeed, the view that scientists typically perform experiments, and that experiments can be considered as ordered collections of steps acting on data and involving a variety of distinct activities, motivates the exploitation of workflow technologies to scientific endeavors. In addition, the scientific domain specifies others requirements as the need to execute in a dynamic environment where resources are not know a priori \cite{DBLP:journals/sigmod/YuB05}. To deal with those requirements, scientific workflows need more flexibility to cope with dynamic environments and more capabilities to manage large data streaming.\\
However, the efforts done for the integration of distributed applications and heterogeneous data are still insufficient. Reusing analytical steps within a workflow instance involves an integration effort. Each analytic step consumes and produces data with a particular structural representation. To compose activities, the structural differences between the activities must be resolved, and this resolution is typically performed by the scientist either manually or by writing a special-purpose program or script \cite{Bowers04}. Moreover, analytical steps used to be rigid and lack of flexibility. That is why, our research focuses on reducing those efforts by providing means, described in section 3, that facilitate the analysis and modeling of scientific applications.\\

The remainder of this paper is organized as follows. Section 2 briefly describes
scientific workflows and motivates its use through an experimental example. Section 3 refers to the flexible cooperative workflow system \textit{Bonita} where we take a closer look to cooperative processes within \textit{Bonita}. In section 4, we defines our proposition for data-intensive science within \textit{Bonita}. In particular, we present a technique for mapping heterogeneous data within scientific workflows based on a binary approach. Section 6 concludes the paper and discusses some future works.

\section{Preliminaries and Motivations}
\label{section:prelimotiv}

Traditionally, the scientific work is concentrated around experiments. Collecting, generating, and analysing large amounts of heterogeneous data is the essence of such work. The scientific experiments which have to be carried out during such a development process are typically composed of series of steps or tasks \cite{medeiros95wasa}. A global abstract view of a scientific work is that it consists of one or more steps with input and output data.\\
Note that there are many others aspects that scientific environments  
has to capture. These include mechanisms for
cooperative work (e.g., ensuring the communication and the coordination for a group to work together in the pursuit of a shared goal) and data management (e.g., managing data format evolution in scientific experiments).
Scientific projects can be considered as ordered collections of 
steps acting on data and involving a variety of distinct
activities. So it motivates the exploitation of the workflow paradigm.\\
Scientific workflow is the application of
workflow technology to scientific endeavours, and is recognized as a
valuable approach for assisting scientists in accessing and analyzing
data. Its applications scopes are the support for large data flows, the need
to decide on further steps after evaluating the previous ones
\cite{medeiros95wasa}, the need to monitor and control workflow execution including ad-hoc changes, and the need to execute in a dynamic environment where resources are not know a priori and may need to adapt to changes \cite{DBLP:journals/sigmod/YuB05}.\\ 
As an example, consider the development of prototypes by digitalization.
We aim to develop, from an original model, a prototype that will be used for testing, measuring and controlling (e.g., metrology) the new model. First, 
we try to identify a number of features belonging to the original model by digitalization it. Then according to the experience, which includes outcomes (measurements, cost aspects, etc.), we conduct another experiments (reconstruction and modification) to simulate it afterwards. If successful, the prototype will be ready to be tested depending on the scientific requirements.\\
From a workflow point of view, this scientific experiment are typically composed of a series of tasks, which are intertwined according to some control sequence (e.g., conditional branches), and receiving/producing manifold information. Thus, the scientific workflow seems to be an appropriate paradigm to integrate it.

\section{Using BONITA for Scientific Workflows}
\label{section:using bonita for scientific workflows}

\begin{figure}[h]
\begin{center}
\scalebox{0.5}{\includegraphics{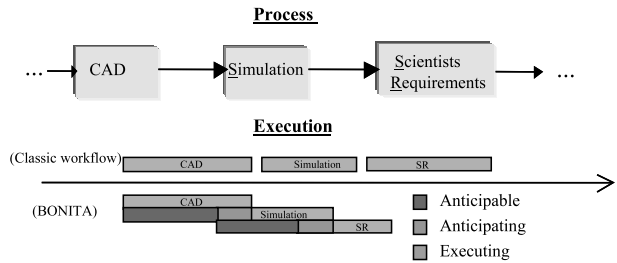}}
\end{center}
\caption{\textbf{Ensuring flexiblity for a scientific workflow execution}}
\label{anticipation}
\end{figure}

Bonita is a flexible cooperative workflow system which incorporates the anticipation of activities as a flexible mechanism of a workflow execution. 
The principle of anticipation is to allow an activity to escape to the traditional start-end synchronization model by ensuring intermediate results to be used as preliminary input into succeeding activities \cite{Grigori01}.
Our approach consists in adding flexibility to scientific workflow executions with minimal changes of the workflow model. We try to reach this goal by relaxing the way the model is interpreted. Scientists can take
some initiatives regarding the way they start the assigned activities.
The use of Bonita will ensure several advantages such as cooperative processes by ensuring numerous interactions between heterogeneous applications,
the communication and the coordination for scientists to work together in the pursuit of a shared goal.\\
As described in section \ref{section:prelimotiv}, we can ensure the flexibility of the scientific experiment. Actually the step of reconstruction and modification, called CAD task, is not completely finished. So we can anticipate it by passing to the next activities with the possibility of return. The anticipation in Bonita offers new alternatives in a dynamic environment. When the user selects an activity, he obtains the list of the executing activities: \textit{executing} or \textit{anticipating} states, 
and his assigned activities: \textit{ready} or \textit{anticipable}
(see figure \ref{anticipation}). Therefore, in our example, by anticipating the CAD's step we will have the opportunity to optimize the process, 
to send a feedback to this step, and eventually add another activity.\\ 

\section{Mapping Data-Intensive Science into BONITA}
\label{section:Mapping Data-Intensive Science into BONITA}

Bonita integrates a great number of services \footnote{Java Open Application Server (JOnAS): J2EETM Platform. http://www.objectweb.org} to control and simplify many cooperative aspects such Java Message Service (JMS) to notify the definition and execution changes within a workflow process, and Jabber 
service that allows the users to receive notifications at real-time and 
exchange different kinds of messages. However, those services lack of heterogeneity and dynamic changes, so they can not deal with the runtime message evolution. Therefore, we need to extend the Workflow engine so that it controls the data flow.\\ 
Storing and transmitting data in binary form is often desirable both
to conserve I/O bandwidth and to reduce storage and processing requirements. PBIO (Portable Binary Input/Output) is a general approach to deal with binary data in storage and transmission \cite{DBLP:journals/tpds/EisenhauerBS02}. Users register the structure of the data that they wish to transmit/store or receive/read and PBIO transparently masks the differences \cite{1069416}. We integrate the PBIO approach to support the messages exchange. It manages the input/output data by storing the streaming of data and matching the appropriate records with the relevant activity. Here, CAD's activity communicates with both Simulation and SR activities where the Simulation activity receives CAD's data as input and SR sends data to CAD as feedback. 

\section{Conclusions and Future Directions}
\label{section:conclusions and future directions}
In this paper, we have presented an approach for scientific workflow
execution. Our approach exploits two distinct specifications for a scientific process: cooperative process and process messaging. We believe that anticipating activities and optimizing data streaming allow dynamic analysis and high performance communication. We plan to further implement this approach by developing a framework adapted for scientific workflow execution  
based on Bonita and the PBIO approach.

\section*{Acknowledgements}
This work was partially supported by Auraryd LLC. Software Company (France), 
within a bilateral cooperation with the ECOO research team (LORIA) on Scientific Workflow and Knowledge Infrastructure in Workflow Management.

\bibliography{WSES06_kgaaloul_charoy_godart}

\end{document}